\documentclass[twocolumn,showpacs,superscriptaddress,aps,prb]{revtex4-2}
\usepackage{bm}
\usepackage{amssymb,amsmath}
\usepackage{graphicx}
\usepackage{color}
\usepackage[colorlinks=true,urlcolor=blue,citecolor=blue]{hyperref}

\begin{document}
\title{Stripe order, impurities, and symmetry breaking in a diluted frustrated magnet}
\author{Xuecheng Ye}
\affiliation{Department of Physics, Missouri University of Science and Technology, Rolla, MO 65409, USA}
\author{Rajesh Narayanan}
\affiliation{Department of Physics, Indian Institute of Technology Madras, Chennai 600036, India.}
\author{Thomas Vojta}
\affiliation{Department of Physics, Missouri University of Science and Technology, Rolla, MO 65409, USA}

\begin{abstract}
We investigate the behavior of the frustrated $J_1$-$J_2$ Ising model on a square lattice
under the influence of random dilution and spatial anisotropies. Spinless impurities generate
a random-field type disorder for the spin-density wave (stripe) order parameter. These random
fields destroy the long-range stripe order in the case of spatially isotropic interactions.
Combining symmetry arguments, percolation theory and large-scale Monte Carlo simulations,
we demonstrate that arbitrarily weak spatial interaction anisotropies restore the stripe phase.
More specifically, the transition temperature $T_c$ into the stripe phase depends
on the interaction anisotropy $\Delta J$ via $T_c \sim 1/|\ln (\Delta J)|$ for small $\Delta J$.
This logarithmic dependence implies that very weak anisotropies are sufficient to restore the
transition temperature
to values comparable to that of the undiluted system. We analyze the critical
behavior of the emerging transition and find it to belong to the disordered two-dimensional Ising
universality class, which features the clean Ising critical exponents and universal logarithmic
corrections. We also discuss the generality of our results and their consequences for experiments.
\end{abstract}

\date{\today}

\maketitle

\section{Introduction}
\label{sec:intro}

The influence of impurities, defects, and other types of quenched random disorder on the symmetry-broken
low-temperature phases of many-particle systems and on their phase transitions is an important topic
in condensed matter physics. Fundamentally, disorder effects are governed by the interplay between
the symmetries of the order parameters characterizing the phase or phase transition and the symmetries
of the disorder (see, e.g., Ref.\ \cite{Vojta19} for a pedagogical discussion).

If the impurities respect the order parameter symmetries, they generically lead to random-$T_c$
disorder, i.e., to spatial variations in the tendency towards the symmetry-broken phase.
As this disorder appears in the mass term of the order parameter field theory,
it is also called random-mass disorder. The diluted ferromagnet is an example for this case
because spinless impurities do not prefer a particular magnetization direction and thus do not
break the spin symmetry. Random-mass disorder can influence phase transitions profoundly, e.g.,
by rounding first-order phase transitions \cite{ImryWortis79,HuiBerker89,AizenmanWehr89}
or by modifying the critical behavior of continuous ones \cite{Harris74}. Quantum phase
transitions can feature additional disorder effects including infinite-randomness critical
points \cite{Fisher92,*Fisher95,MMHF00,HoyosKotabageVojta07,*VojtaKotabageHoyos09},
smeared phase transitions \cite{Vojta03a,*HoyosVojta08},
and quantum Griffiths singularities \cite{Griffiths69,ThillHuse95,RiegerYoung96,*YoungRieger96}
(see Refs.\ \cite{Vojta06,*Vojta10,Vojta13} for reviews).

If, on the other hand, the impurities locally break the order parameter symmetries, a stronger
coupling between the disorder and the order parameter can be expected. The generic result is
random-field disorder \cite{ImryMa75}, i.e., randomness in the field conjugate to the order parameter in the
corresponding field theory. More complicated scenarios such as random-easy-axis disorder
\cite{HarrisPlischkeZuckermann73,PelcovitsPytteRudnick78,Fisher85,DudkaFolkHolovatch05,Mirandaetal21}
can occur if the impurities break the order parameter symmetries only partially.
Random fields can have more dramatic effects than random-mass disorder.
In sufficiently low space dimensions ($d \le 2$ for discrete order parameter symmetry and $d\le 4$
for continuous order parameter symmetry), even weak random fields
destroy the symmetry-broken phase itself via domain formation \cite{ImryMa75,Binder83,AizenmanWehr89}.

Recent years have seen renewed interest in phases that spontaneously break real-space
symmetries in addition to spin, phase, or gauge symmetries, including the charge-density wave or
stripe phases in cuprate superconductors \cite{EmeryKivelsonTranquada99,KBFOTKH03,VojtaM09},
the Ising-nematic phases in the iron
pnictides \cite{FKLEM10,FernandesChubukovSchmalian14,FradkinKivelsonTranquada15},
as well as valence-bond solids in certain quantum magnets
\cite{ReadSachdev89,MambriniLauchliPoilblancMila06,Sandvik07}. In general, impurities locally
break the real-space symmetries of the associated order parameters. They thus generically lead
to random-field type disorder for such order parameters
\cite{Fernandez88,FyodorovShender91,CDFK06,LohCarlsonDahmen10,NieTarjusKivelson14,KunwarSenVojtaNarayanan18,AndradeHoyosRachelVojta18,Mirandaetal21}.
In addition to destroying the original long-range order, these random fields can also induced novel
phases of matter \cite{AndradeHoyosRachelVojta18,Mirandaetal21}.

A prototypical model for impurity-induced random fields is the frustrated $J_1$-$J_2$ Ising model on a
square lattice, with ferromagnetic nearest-neighbor interactions and antiferromagnetic
next-nearest-neighbor interactions. For sufficiently strong next-nearest-neighbor interactions,
it features a stripe-ordered low-temperature phase. As site or bond dilution locally break
the symmetry between the two equivalent stripe directions, they generate random fields
for the nematic order \cite{Fernandez88,KunwarSenVojtaNarayanan18} which destroy the
stripe phase via domain formation. Interestingly, the strength of the random fields can
be tuned by the repulsion between the impurities \cite{KunwarSenVojtaNarayanan18}.

In the present paper, we revisit the diluted $J_1$-$J_2$ Ising model and focus on the
interplay between the random-field  disorder and global interaction anisotropies that may arise,
e.g., from strain engineering, epitaxial growth or the shape of crystallites or samples.
We combine symmetry arguments, percolation theory and large-scale Monte Carlo simulations
to show that the stripe phase is restored by an arbitrarily weak global anisotropy
(modeled, e.g., by a difference $\Delta J$ between the horizontal and vertical interaction strengths)
that explicitly breaks the symmetry between the two stripe directions. Importantly,
the transition temperature $T_c$ into the stripe phase varies with the interaction
anisotropy as $T_c \sim 1/|\ln(\Delta J)|$. This logarithmic dependence
implies that a very weak anisotropy is sufficient to suppress most random-field effects
and restore the transition temperature to a value comparable to that of the undiluted system.
We also determine the critical behavior of the emerging phase transition between the paramagnetic
and stripe phases.
Just as the transition in the diluted Ising ferromagnet, it belongs to the disordered
two-dimensional Ising universality class which is characterized by the clean Ising
exponents and universal logarithmic corrections.

The remainder of our paper is organized as follows. In Sec.\ \ref{sec:model+theory}, we define the
$J_1$-$J_2$ Ising model. We also discuss the random-field mechanism and domain formation.
Our computer simulation methods are introduced in Sec.\ \ref{sec:MC}. Section \ref{sec:results}
is devoted to the simulation results and a comparison with theoretical predictions. We conclude in
Sec.\ \ref{sec:conclusion} by discussing the generality of our findings and their consequences for
experiments.

\section{Model and random-field mechanism}
\label{sec:model+theory}
\subsection{Diluted anisotropic $J_1$-$J_2$ Ising model}
\label{subsec:model}

We start with the well-known $J_1$-$J_2$ Ising model on a square lattice of $N=L^2$ sites given
by the Hamiltonian
\begin{equation}
H_0 = - J_1 \sum_{\langle ij \rangle} S_i S_j  - J_2 \sum_{\langle\langle ij \rangle\rangle}  S_i S_j ~.
\label{eq:H0}
\end{equation}
Here, $S_i = \pm 1$ is a classical Ising spin, $\langle ij \rangle$ denotes pairs of nearest-neighbor
sites coupled by the ferromagnetic interaction $J_1>0$, and $\langle\langle ij \rangle\rangle$
denotes next-nearest neighbor pairs coupled by the antiferromagnetic interaction $J_2 < 0$.
The phases of this system are well-understood (see, e.g., Refs.\
\cite{JinSenSandvik12,JinSenGuoSandvik13,KalzHoneckerMoliner11,KalzHonecker12} and references therein).
It displays paramagnetic behavior at high temperatures.  As the temperature is lowered, two distinct
long-range ordered phases appear.
For $|J_2|/J_1 < 1/2$, the low-temperature phase is ferromagnetic; it breaks the $Z_2$ Ising
spin symmetry but none of the real-space symmetries.
For $|J_2|/J_1>1/2$, in contrast, the low-temperature phase features a stripe-like spin order
that breaks not only the Ising spin symmetry but also the $C_4$ rotation symmetry of the square lattice.

\begin{figure}
\includegraphics[width=2.6cm]{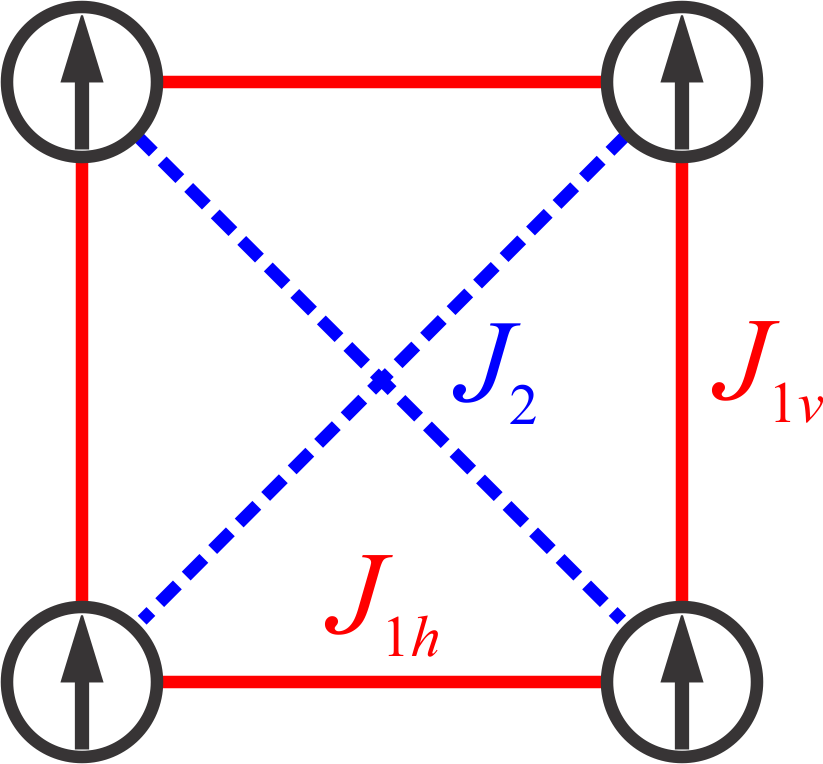}
\caption{Interactions of the anisotropic $J_1$-$J_2$ model.}
\label{fig:lattice}
\end{figure}

To explore the combined influence of quenched disorder and spatial anisotropies on the
stripe phase, we now introduce site dilution, and we allow the nearest-neighbor interaction
to take different values $J_{1h}$ and $J_{1v}$ for horizontal and vertical bonds, respectively
(see Fig.\  \ref{fig:lattice}).
The resulting Hamiltonian reads
\begin{eqnarray}
H =&& - J_{1h} \sum_{\langle ij \rangle_h}\epsilon_i\epsilon_j S_i S_j - J_{1v} \sum_{\langle ij \rangle_v}\epsilon_i\epsilon_j S_i S_j \nonumber \\
     && - J_2 \sum_{\langle\langle ij \rangle\rangle}\epsilon_i\epsilon_j  S_i S_j ~.
\label{eq:H}
\end{eqnarray}
The $\epsilon_i$ are quenched random variables that can take the values 0 (representing a vacancy)
with probability $p$ and 1 (occupied site) with probability $1-p$. We consider the $\epsilon_i$
at different sites statistically independent; the effects of (anti)correlations between the
vacancies were explored in Ref.\ \cite{KunwarSenVojtaNarayanan18}. We parameterize the
nearest-neighbor interactions in terms of their average and difference,
$J_{1h} = J_1 + \Delta J,~ J_{1v} = J_1 - \Delta J$. In the following, we focus on the parameter
region that favors stripe order at low temperatures, i.e., on $|J_2|/J_1 > 1/2$.

\subsection{Random-field disorder}
\label{subsec:rf}

\begin{figure}
\includegraphics[width=\columnwidth]{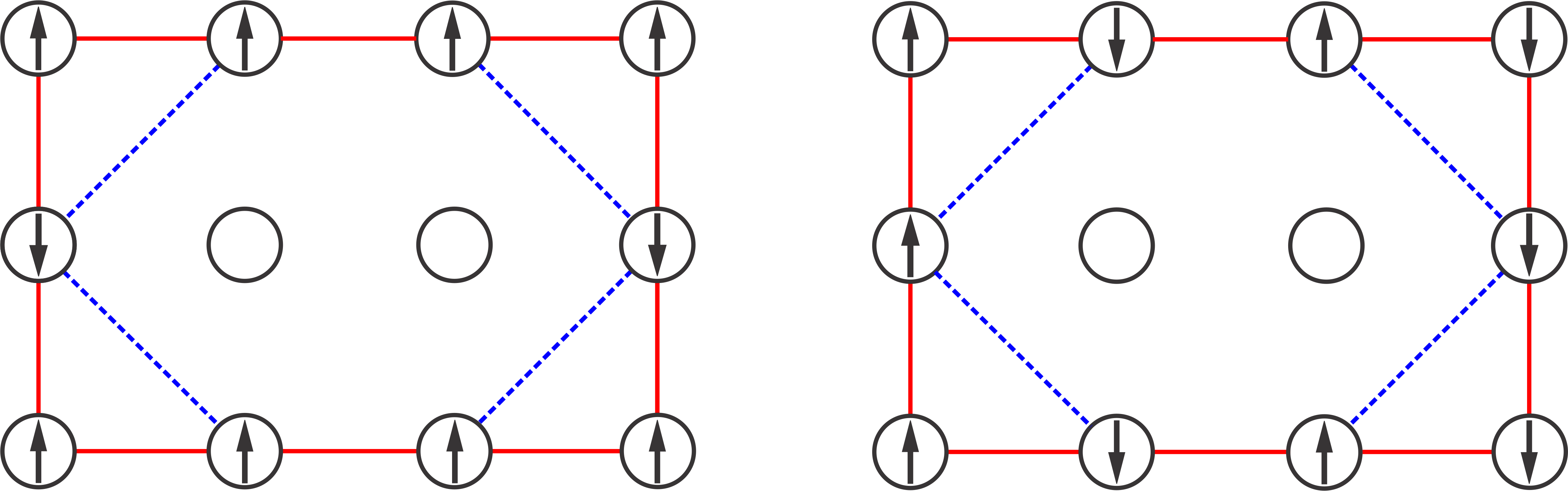}
\caption{Random-field mechanism: A pair of vacancies on horizontal nearest-neighbor sites prefers
horizontal stripes (left) over vertical stripes (right) by an energy difference of $2 J_1$.}
\label{fig:vacancy_pair}
\end{figure}

While a single vacancy does not break the $C_4$ rotation symmetry of the lattice, spatial arrangements
of several vacancies generally do break this symmetry locally, leading to the emergence of random-field
disorder that locally prefers one stripe direction over the other (even in the absence of
interaction anisotropies, i.e., for $\Delta J=0$). Specifically, a pair of vacancies on
horizontal nearest-neighbor sites prefers
horizontal stripes over vertical stripes by an energy difference of $2 J_1$,
see Fig.\ \ref{fig:vacancy_pair} \cite{Fernandez88,KunwarSenVojtaNarayanan18}.
Analogously, a vacancy pair on vertical nearest-neighbor sites prefers vertical stripes.

\begin{figure}
\includegraphics[width=\columnwidth]{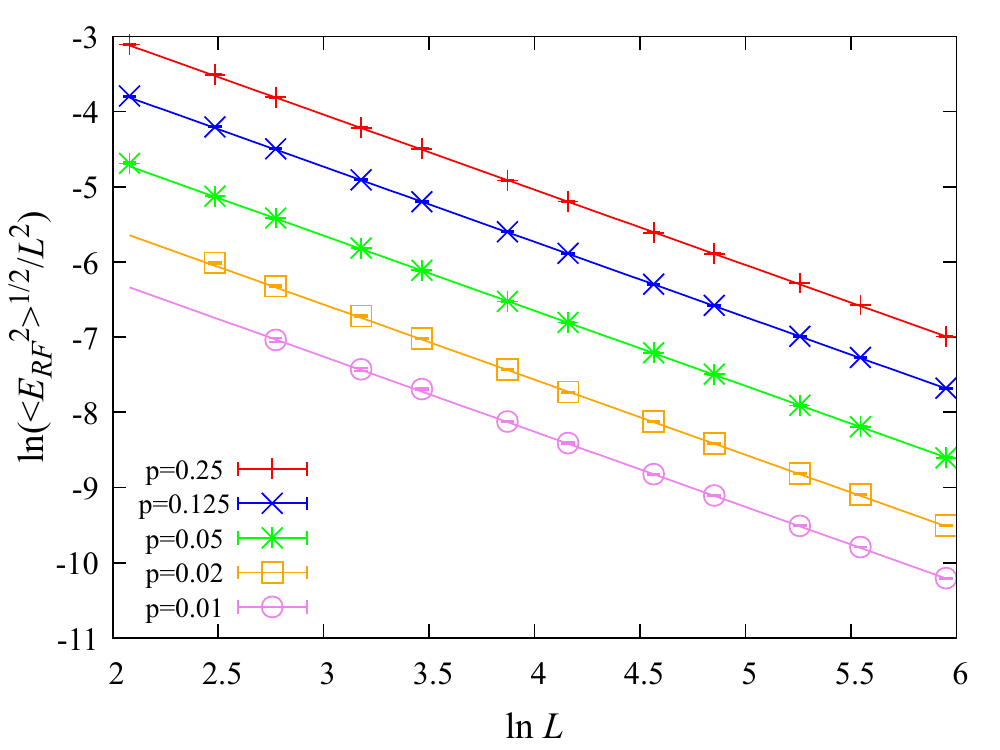}
\caption{Root-mean-square random-field energy of a perfect stripe state per lattice site, $\langle E_{RF}^2\rangle^{1/2}/L^2$,
vs.\ linear system size $L$ for several dilutions $p$. The data are determined by averaging the square of the energy difference
between perfect horizontal and vertical stripe states over 20,000 disorder configurations.
The solid lines represent relation (\ref{eq:E_RF}) without adjustable parameters. }
\label{fig:E_RF}
\end{figure}

The typical random-field energy of a perfect (horizontal or vertical) stripe state
in a system of $L\times L$ sites can be easily estimated in the
limit of low dilution $p$ when different vacancy pairs can be considered independent and
arrangements of three or more vacancies on neighboring sites are suppressed.
A system of $L\times L$ sites has $2L^2$ distinct nearest neighbor pairs (bonds), resulting
in an average number of vacancy pairs of $2 L^2 p^2$. The random-field energy $E_{RF}(L)$
is thus the sum of $2 L^2 p^2$ random contributions $\pm J_1$. The central limit theorem
then gives
\begin{equation}
\langle E_{RF}^2(L) \rangle = 2 L^2 p^2 J_1^2 = h_\mathrm{eff}^2 L^2
\label{eq:E_RF}
\end{equation}
with effective random field strength $h_\mathrm{eff} = \sqrt{2} p J_1$
\footnote{Note that the effective random-field strength is proportional to $p$
rather than $p^2$ as one might have naively expected because the probability
for finding a vacancy pair is proportional to $p^2$.}.
We have confirmed the relation (\ref{eq:E_RF}) numerically for a range of dilutions
and system sizes, as can be seen in Fig.\ \ref{fig:E_RF}.
It holds (at least in very good approximation) for dilutions as high as $p=1/4$.

\subsection{Domain formation}
\label{subsec:domain}

According to Imry and Ma \cite{ImryMa75}, the fate of the symmetry-broken low-temperature phase
is governed by the competition between the random-field energy gain due to the formation
of domains of horizontal and vertical stripes that align with the local random field
and the energy cost of a domain wall. The energy cost of a straight domain wall between
horizontal and vertical stripes in the undiluted $J_1$-$J_2$ model is easily worked out,
it equals $2 |J_2|$ per
lattice constant. This domain formation problem can be mapped onto a random-field Ising
model with the Ising variable representing the difference between horizontal and vertical
stripes in the $J_1$-$J_2$ model (\ref{eq:H}).

Let us first consider the case of isotropic interactions, $\Delta J = 0$ (which maps onto an unbiased
random-field Ising model). In two dimensions, domains appear for arbitrarily weak random fields
beyond the so-called breakup length scale $L_0$. For weak random fields, $L_0$ depends
exponentially on the ratio between the domain wall energy scale $J_2$ and the random-field strength
$h_\mathrm{eff}$,
\begin{equation}
L_0 = A \exp(c J_2^2/h_\mathrm{eff}^2)
\label{eq:L0}
\end{equation}
with $A$ and $c$ constants \cite{Binder83}. As horizontal and vertical stripe domains are equally likely for $\Delta J =0$,
the domain formation destroys the symmetry-broken low-temperature phase.
(A rigorous proof that the Gibbs state in a two-dimensional random-field Ising model
is unique was given by Aizenman and Wehr \cite{AizenmanWehr89}.) This agrees with the
Monte Carlo simulation results of Ref.\ \cite{KunwarSenVojtaNarayanan18}.

For anisotropic interactions, $\Delta J \ne 0$, the problem maps onto a biased random-field
Ising model. In the case $\Delta J >0$, horizontal stripes are preferred over vertical
ones. Minority (vertical stripe) domains have a finite maximum size that decreases with
increasing $\Delta J$ \cite{Binder83}. At low temperatures, we thus expect the system to
consist of finite-size vertical-stripe domains embedded in the bulk featuring horizontal
stripes.

The domains of the two-dimensional random-field Ising model were further
investigated by Sepp\"al\"a et al.\ \cite{SeppalaPetejaAlava98,*SeppalaAlava01} and by
Stevenson and Weigel \cite{StevensonWeigel11,*StevensonWeigel11b}. They demonstrated that
the domain structure in the unbiased case on length scales larger than $L_0$
resembles the fractal cluster structure of a critical percolation problem, at least for
sufficiently weak random fields (i.e., sufficiently large $L_0$). Increasing bias
($\Delta J >0$) drives the domain pattern away from percolation criticality, and
a massive spanning cluster of the majority stripes forms. This transition in the
domain structure is governed by the usual two-dimensional classical percolation exponents.

\subsection{Magnetic phase transition}
\label{subsec:transition}

The random-field disorder in the diluted $J_1$-$J_2$ model locally breaks the $C_4$ rotation
symmetry of the square lattice. However, it does not break the $Z_2$ Ising spin symmetry.
This leaves open the possibility of a magnetic phase transition into a long-range ordered
low-temperature phase that spontaneously breaks this remaining $Z_2$ symmetry
\footnote{Note that situation differs from the random-field Ising model where the random fields
completely break the order parameter symmetry.}. This phase
transition, if any, has to occur on the background of the stripe domain pattern discussed
in Sec.\ \ref{subsec:domain}.

In the absence of a global anisotropy (i.e., for $\Delta J = 0$), the magnetic phase transition
is impossible because the domain structure resembles critical percolation. This implies that
neither horizontal nor vertical domains form a massive cluster that covers a
finite fraction of the lattice sites and can support long-range magnetic order. This conclusion
agrees with the Monte Carlo results of Ref.\ \cite{KunwarSenVojtaNarayanan18}.

In the presence of a global anisotropy, in contrast, the majority stripes
(horizontal stripes for $\Delta J > 0$) form a massive infinite (spanning) cluster. The
Ising spins on this cluster can therefore spontaneously break the $Z_2$ Ising symmetry
and develop long-range order. To estimate the critical temperature $T_c$ of the magnetic transition
as function of the global anisotropy $\Delta J$, we recall that the critical temperature
of a diluted Ising model close to the percolation threshold $p_c$ varies as $T_c \sim 1/|\ln(p-p_c)|$
with the distance $p-p_c$ from the threshold (see, e.g., \cite{StaufferAharony_book91,Cardy_book96}).
In our $J_1$-$J_2$ model (\ref{eq:H}), the distance of the stripe domain pattern from percolation
criticality is controlled by $\Delta J$. We therefore expect the transition temperature into the
stripe phase to vary as
\begin{equation}
T_c \sim 1/|\ln(\mathrm{const} \, \Delta J)|~.
\label{eq:Tc}
\end{equation}

In addition to random-field disorder, the vacancies also create random-mass disorder which is
known to prevent first-order phase transitions in two dimensions \cite{ImryWortis79,HuiBerker89,AizenmanWehr89}.
We thus expect the transition into the stripe phase to be continuous. On symmetry grounds, its critical behavior
should belong to the two-dimensional
disordered Ising universality class as it spontaneously breaks the remaining $Z_2$ symmetry.
This is a particularly interesting universality
class because the clean two-dimensional Ising correlation length exponent takes the value $\nu=1$ which
makes it marginal with respect to the Harris criterion \cite{Harris74} $d\nu > 2$.
Perturbative renormalization-group studies \cite{DotsenkoDotsenko83,Shalaev84,Shankar87}
predict that the critical behavior of the disordered Ising model is controlled by the clean
Ising fixed point. Disorder, which is a marginally irrelevant operator, gives rise to universal
logarithmic corrections to scaling. Early computer simulations \cite{FahnleHoleyEckert92,KimPatrascioiu94,Kuhn94},
in contrast, found nonuniversal critical exponents that vary continuously with disorder strength.
More recent large-scale simulations strongly support the logarithmic-corrections scenario
(see Ref.\ \cite{ZWNHV15} and references therein).

\section{Monte Carlo simulations}
\label{sec:MC}

In order to gain a quantitative understanding of the interplay between the random fields and the
global anisotropy in the $J_1$-$J_2$ model, we perform extensive Monte Carlo simulations of the
Hamiltonian (\ref{eq:H}). As we are interested in the fate of the stripe low-temperature
phase, we fix the interaction energies at the values $J_1=-J_2=1$ for which
the undiluted isotropic system enters the stripe phase at a temperature of about 2.08
\cite{KalzHoneckerMoliner11}.
The dilution is fixed at $p=0.25$. This relatively strong disorder leads to moderate domain
sizes that actually fit into the sample sizes we are able to simulate. The global
interaction anisotropy $\Delta J$ is varied between 0 and 0.2.

In the parameter region $J_1 > 0, J_2 < 0$, the interactions of the $J_1$-$J_2$ model are
frustrated. Therefore, cluster algorithms such as the Wolff \cite{Wolff89} and
Swendsen-Wang \cite{SwendsenWang87} algorithms
do not improve the efficiency of the simulations \cite{KalzHoneckerFuchsPruschke08}.
We therefore combine conventional single-spin-flip Metropolis updates \cite{MRRT53}
with ``corner'' updates that exchange the two spins on the diagonal corners of a
$2\times 2$ plaquette of sites. These corner updates locally turn horizontal stripes
into vertical ones and vice versa. Specifically, a full Monte Carlo sweep consists
of a Metropolis sweep over the full lattice followed by two corner sweeps (one attempting to
exchange the top right and bottom left sites of each plaquette, the other doing the same
for the top left and bottom right sites).

\begin{figure}
\includegraphics[width=\columnwidth]{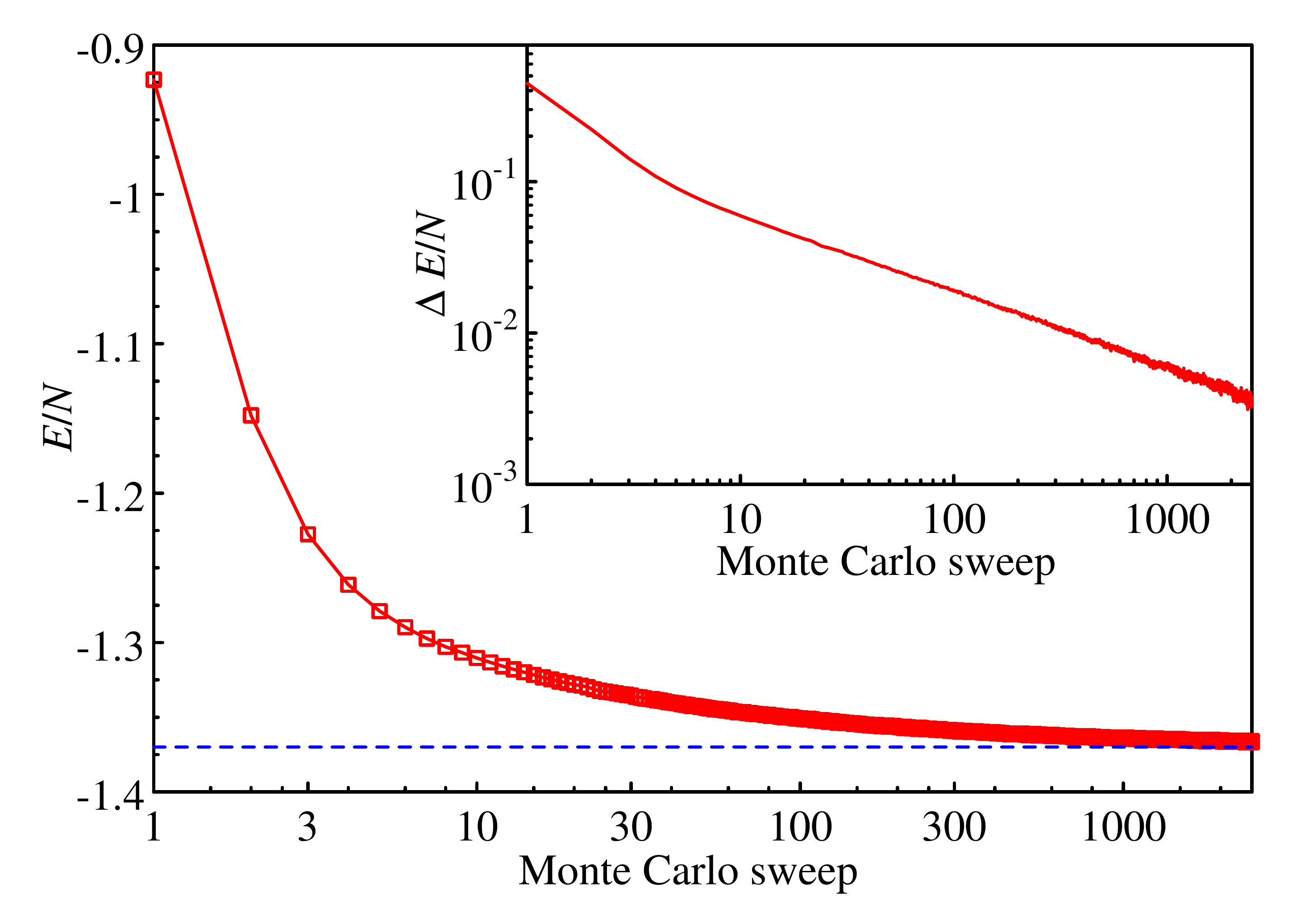}
\caption{Energy per site $E/N$ vs.\ Monte Carlo sweep for a system of linear size $L=96$,
$\Delta J =0$ and temperature $T=1.15$. The data are averages over 3000 runs, each with a
different disorder configuration. The simulations start from a random configuration of spins
(hot start). The dashed line marks the equilibrium value of $E/N$.
Inset: Log-log plot of the deviation $\Delta E$ from the equilibrium value vs.\ Monte Carlo
sweep.}
\label{fig:energy_equilibration}
\end{figure}

As both Monte Carlo moves are local, equilibration is slow, and the problem is further
exacerbated by the random-field effects at nonzero dilution. This is illustrated in Fig.\
\ref{fig:energy_equilibration} which shows how the energy approaches its equilibrium value
(for a prototypical set of parameters).
The data demonstrate that the relaxation is slower than exponential, it approximately follows a
power law over at least two orders of magnitude in Monte Carlo time.

\begin{figure}
\includegraphics[width=\columnwidth]{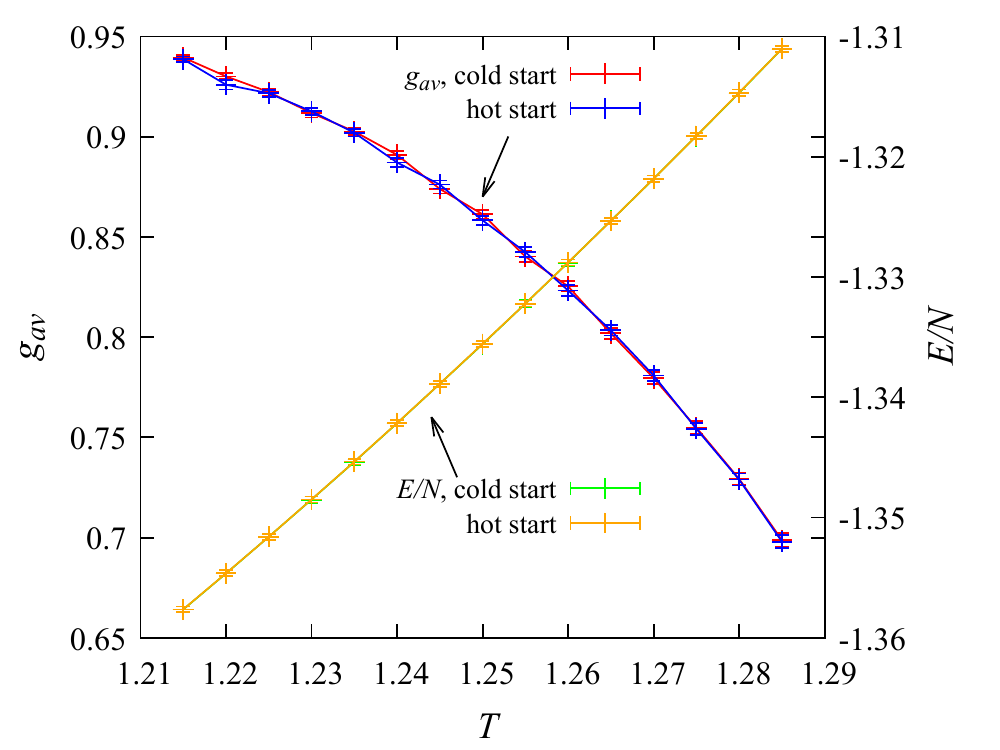}
\caption{Comparison of simulations with hot starts (random initial spin configuration, run starts
at highest temperature) and cold starts (spins initially in perfect stripe state, run starts
at lowest temperature). Shown are the average Binder cumulant $g_\mathrm{av}$ and the total energy per site $E/N$
as function of temperature $T$ for a system with $L=96$, $\Delta J =0.01$.
The data are averages over 5000 runs, each with a different disorder configuration,
using $3\times 10^5$ equilibration sweeps and $4 \times 10^5$ measurement sweeps. }
\label{fig:hot_cold}
\end{figure}

Consequently, long equilibration periods are required in the simulations, as well as long
measurement periods to ensure that the measurements do not remain correlated over the simulation
run. This severely limits the system sizes we can study.
We employ equilibration periods ranging from 30,000 full sweeps for the smallest
systems (linear size $L=16$) to $10^6$ sweeps for the largest systems studied ($L=192$).
The corresponding measurement periods range from 30,000 to $2\times 10^6$ full sweeps, with
a measurement taken after each sweep. We also change the temperature in small steps and
use the final spin configuration for one temperature as the initial configuration for the next.
To check whether the observables truly reach their equilibrium values (within the statistical
errors), we compare the results of runs with ``hot'' starts (spins have independent random
values initially) and ``cold'' starts (spins are in perfect stripe state initially).
An example of such a comparison is shown in Fig.\ \ref{fig:hot_cold}.
All data are averaged over 3,000 to 100,000 disorder (vacancy) configurations, depending
on system size and temperature range.

During the simulations, we compute a number of observables including the total energy per site
$[\langle e \rangle]_\mathrm{dis}$ and the specific heat
$C = (N/T^2)[\langle e^2 \rangle - \langle e \rangle^2]_\mathrm{dis}$. Here, $e=E/N$ stands for an
individual energy measurement, $\langle \ldots \rangle$ is the canonical thermodynamic average
(which is approximated by the Monte Carlo average) and $[ \ldots]_\mathrm{dis}$ is the average
over the disorder configurations. We also calculate the two-component stripe order parameter
$\psi=(\psi_h,\psi_v)$ with
\begin{equation}
\psi_h = \frac 1 N \sum_i (-1)^{y_i} \epsilon_i S_i ~, \quad\psi_v = \frac 1 N \sum_i (-1)^{x_i} \epsilon_i S_i ~.
\label{eq:stripe_OP}
\end{equation}
Here, the indices $h$ and $v$ denote horizontal and vertical stripe order, respectively, and
$x_i$ and $y_i$ are the (integer) coordinates of site $i$. The corresponding stripe susceptibility reads
$\chi_s =(N/T)[\langle |\psi|^2 \rangle - \langle |\psi| \rangle^2]_\mathrm{dis}$.
Dimensionless observables are particularly useful for finding the phase transition temperature
and analyzing the critical behavior. We therefore also determine the average and global Binder cumulants
\begin{equation}
g_\mathrm{av} = \left[ 2 - \frac{\langle |\psi|^4 \rangle}{\langle |\psi|^2 \rangle^2}\right]_\mathrm{dis}~, \quad
g_\mathrm{gl} =  2 - \frac{[\langle |\psi|^4 \rangle]_\mathrm{dis}}{[\langle |\psi|^2 \rangle]_\mathrm{dis}^2}
\label{eq:Binder}
\end{equation}
With increasing system size, these Binder cumulants are expected to approach the values 0 in the disordered phase
and 1 in the stripe-ordered phase, and curves of the Binder cumulants vs.\ temperature for different system sizes
cross at the phase transition temperature.  $g_\mathrm{av}$ and $g_\mathrm{gl}$ capture similar information and are
expected to have identical scaling behaviors, but they differ in how the
disorder average is performed. For the average Binder cumulant $g_\mathrm{av}$, an individual Binder cumulant is
computed for each disorder configuration. These individual values are then averaged to yield $g_\mathrm{av}$.
To obtain the global Binder cumulant $g_\mathrm{gl}$, in contrast, the second and fourth moment of the stripe
order parameter are averaged over the disorder configurations, and the cumulant is then constructed
from these disorder-averaged values. In the present paper, we employ the average Binder cumulant for most
of the analysis because it shows weaker corrections to scaling at the transition into the stripe phase.

\section{Results}
\label{sec:results}
\subsection{Isotropic interactions, $\Delta J =0$}
\label{subsec:res_isotropic}

To test our simulation and data analysis techniques, we first consider $\Delta J = 0$, i.e.,
equal exchange interactions $J_{1h}$ and $J_{1v}$ in the horizontal and vertical directions,
respectively. This case can be compared with Ref.\ \cite{KunwarSenVojtaNarayanan18} and
serves as the reference case for studying the effects of anisotropic interactions.

\begin{figure}
\includegraphics[width=\columnwidth]{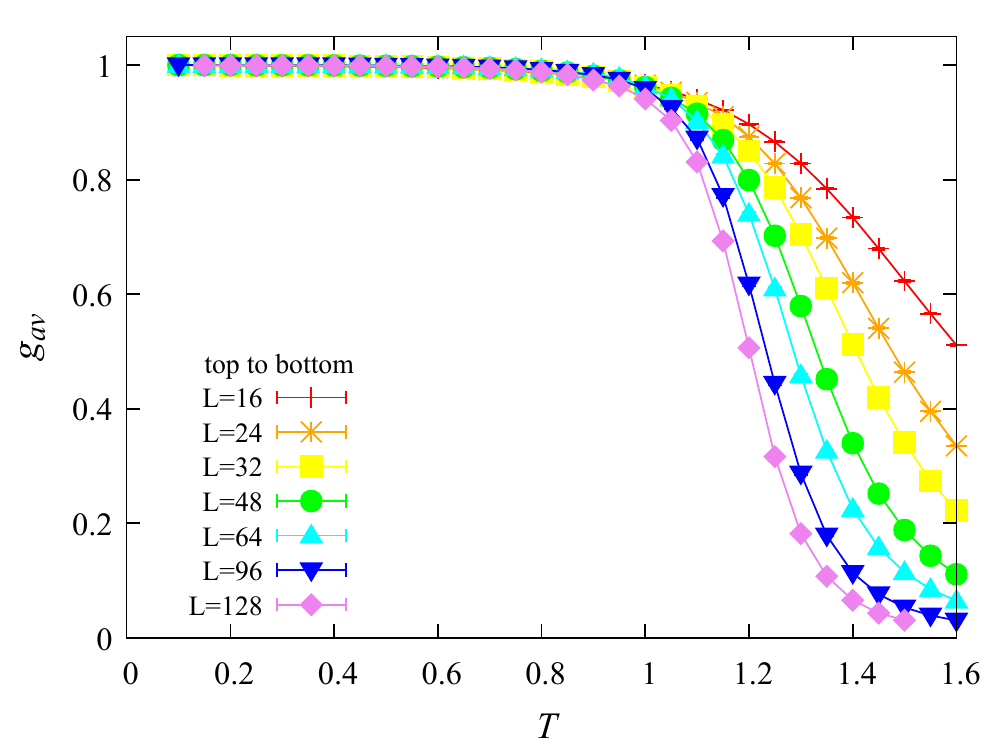}
\caption{Average Binder cumulant $g_\mathrm{av}$ vs.\ temperature $T$ for isotropic
interactions $\Delta J=0$ and several system sizes $L$. $p=1/4$, $J_1=-J_2=1$.
The data are averages over 3000 to 5000 disorder configurations. The resulting statistical
errors are smaller than the symbol size.}
\label{fig:00_gav}
\end{figure}

Figure \ref{fig:00_gav} presents the Monte Carlo simulation results for the average stripe
Binder cumulant $g_\mathrm{av}$ as a function of temperature $T$ for several system sizes $L$
at dilution $p=1/4$ and  $J_1=-J_2=1$.
The curves for different $L$ do not cross, instead $g_\mathrm{av}$ approaches zero with
increasing $L$. The global Binder cumulant $g_\mathrm{gl}$ behaves analogously
\footnote{Note that $g_\mathrm{gl}$ and $g_\mathrm{av}$ have different low temperature limits.
$g_\mathrm{av}$ approaches unity because the Binder cumulant of every single disorder realization
reaches this value (assuming the ground state is unique up to symmetries). $g_\mathrm{gl}$,
in contrast, may approach a value below unity if different disorder realizations feature different
order parameter values for $T \to 0$.}.
This implies that there is no phase transition, and the system does not enter
a long-range ordered stripe phase. This agrees with the expectation of domain formation
according to the Imry-Ma argument discussed in Sec.\
\ref{subsec:domain} and with the results of Ref.\ \cite{KunwarSenVojtaNarayanan18}.

\begin{figure*}
\includegraphics[height=4.2cm]{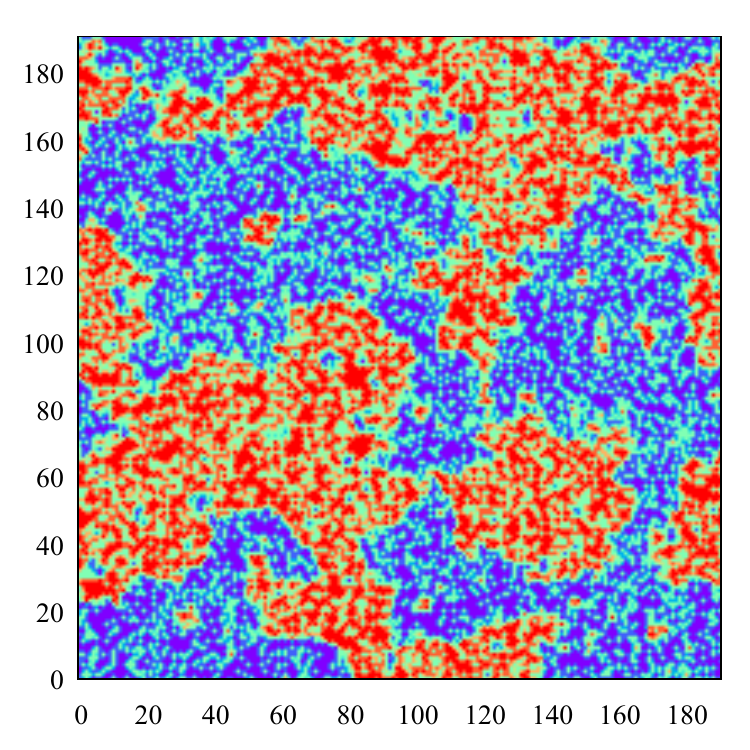}
\includegraphics[height=4.2cm]{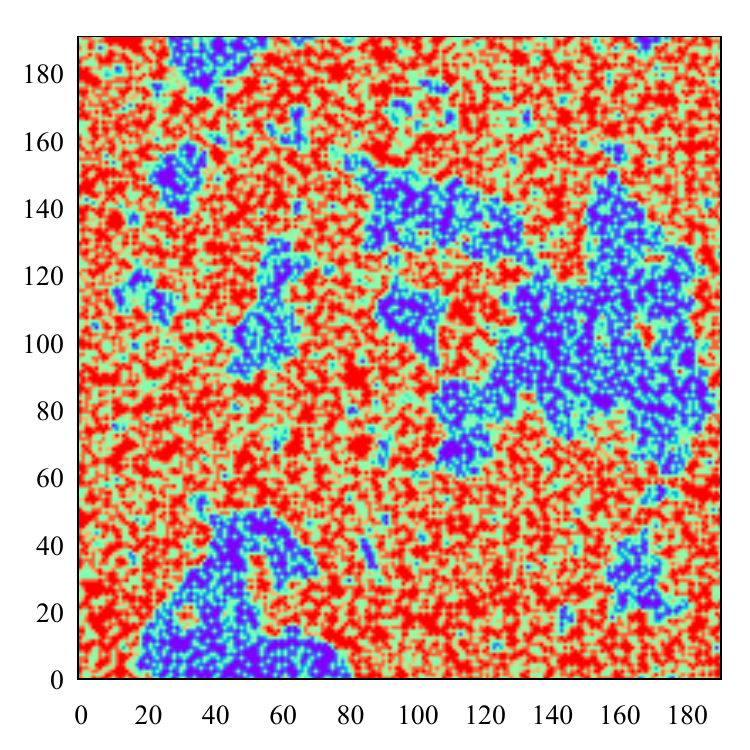}
\includegraphics[height=4.2cm]{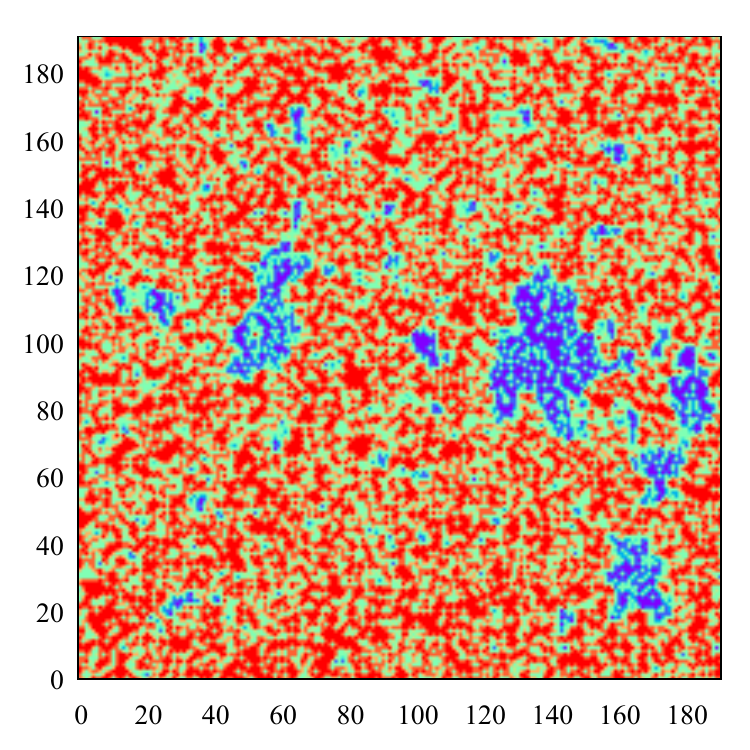}
\includegraphics[height=4.2cm]{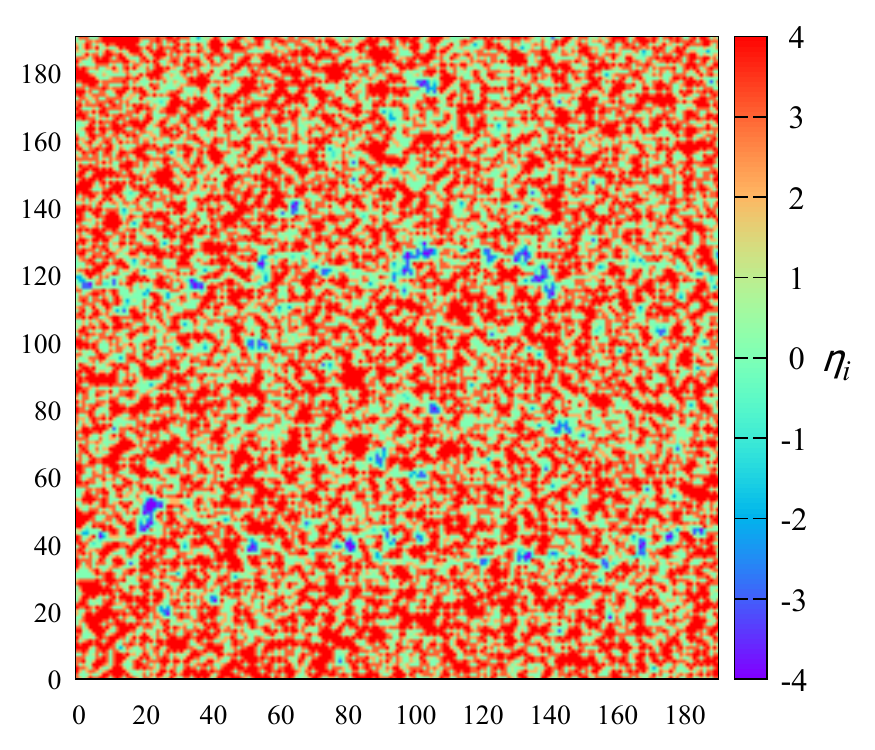}
\caption{Snapshots of the local nematic order parameter $\eta_i$ of one particular disorder configuration
for several anisotropies: $\Delta J = 0$, 0.002, 0.01, 0.05 (left to right).
The data are taken a temperature $T=0.1$ reached via
simulated annealing from high temperatures. $L=192$, $p=1/4$, $J_1=-J_2=1$. }
\label{fig:heatmaps_nem}
\end{figure*}

The domains can be seen explicitly in a snapshot of the local nematic order parameter $\eta_i$
in Fig.\ \ref{fig:heatmaps_nem}.
It is defined via a sum over all bonds from site $i$ to its nearest neighbors,
$\eta_i = \sum'_{j} \epsilon_i \epsilon_j S_i S_j f_{ij}$ where $f_{ij}=1$ for horizontal bonds
and $-1$ for vertical bonds. (This means that $\eta_i=4$ for perfect horizontal stripe order
and $-4$ for perfect vertical stripe order.)

The figure indicates that horizontal and vertical stripes are equally likely for  $\Delta J =0$, as expected in
the isotropic case. It also suggests a breakup length $L_0$ in the range between
about 50 and 100 lattice constants.
It is interesting to compare this estimate with the random-field Ising model result (\ref{eq:L0}).
Using the values $A \approx 6.1$ and $c \approx 1.9$ found numerically by Sepp\"al\"a et al.\
\cite{SeppalaPetejaAlava98,*SeppalaAlava01}, eq.\ (\ref{eq:L0}) yields a breakup length of
about $2 \times 10^7$ for $p=1/4$, much larger than the length identified in Fig.\ \ref{fig:heatmaps_nem}.
We believe that this stems from the fact that the domain wall energy in the diluted system
is significantly smaller than the value $2 |J_2|$ per unit cell in the undiluted system because
the domain wall can make use of the vacancies to reduce the number of unfulfilled bonds. In fact, assuming
that the vacancies reduce the domain wall energy by a factor of 2 to 3,  eq.\ (\ref{eq:L0}) yields
breakup length values comparable to the sizes seen in Fig.\ \ref{fig:heatmaps_nem}.

Thus, the vacancies play a complex role in the destruction  of the stripe order: They generate random
fields, they renormalize the domain wall energy, and they create random-mass disorder.

\subsection{Anisotropic interactions, $\Delta J >0$}
\label{subsec:res_anisotropic}

We now turn to the main topic of this manuscript, the effects of a weak global interaction
anisotropy  $\Delta J$. To this end, we perform Monte Carlo simulations for
$\Delta J = 0.002$, 0.005, 0.01, 0.02, 0.05, 0.1, and 0.2. Snapshots of the resulting local nematic
order parameter $\eta_i$ at low temperatures are presented in Fig.\ \ref{fig:heatmaps_nem}
for a few characteristic $\Delta J$ values. As expected from the discussion in Sec.\
\ref{subsec:domain}, the snapshots show that horizontal stripes proliferate with increasing
$\Delta J$ and form an infinite spanning cluster while vertical stripes are restricted to
finite-size clusters. Already at $\Delta J = 0.05$, vertical stripe domains have essentially
vanished.

\begin{figure}
\includegraphics[width=\columnwidth]{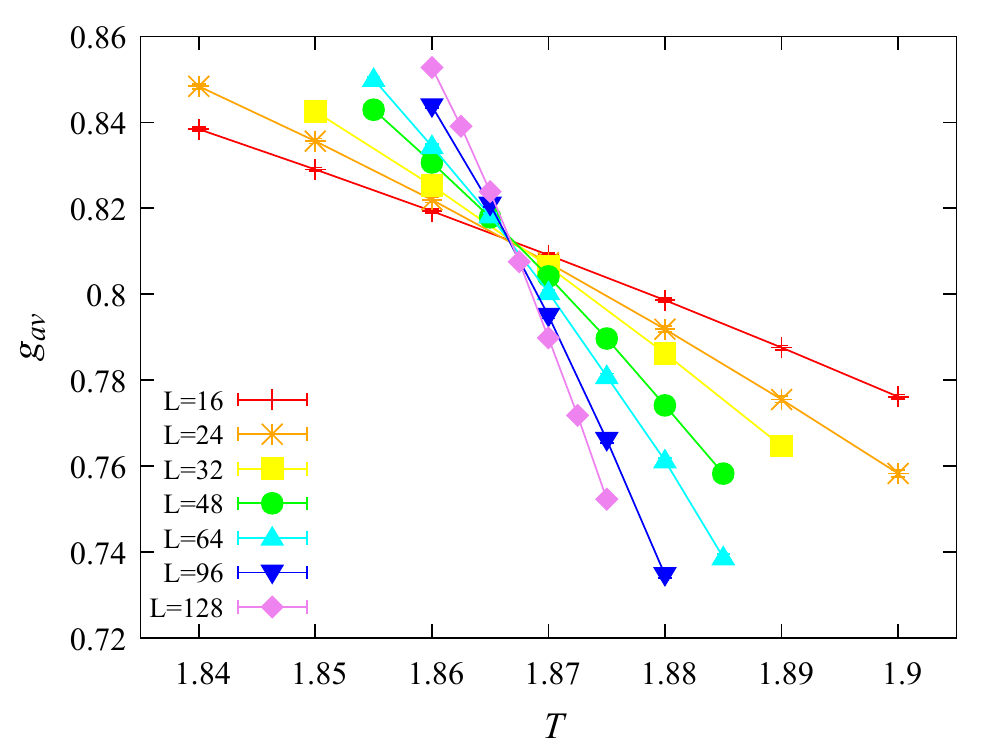}
\caption{Average Binder cumulant $g_\mathrm{av}$ vs.\ temperature $T$ for anisotropic
interactions with $\Delta J=0.2$ and several system sizes $L$. $p=1/4$, $J_1=-J_2=1$.
The data are averages over 30,000 to 100,000 disorder configurations. The resulting statistical
errors are much smaller than the symbol size.}
\label{fig:02_gav}
\end{figure}

\begin{figure}
\includegraphics[width=\columnwidth]{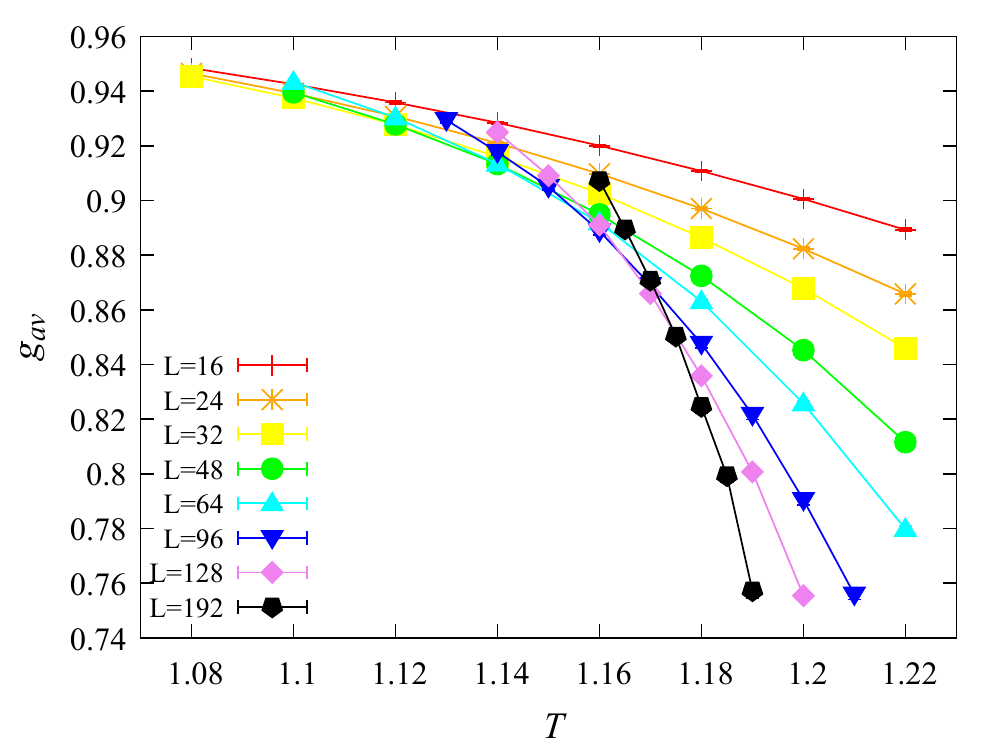}
\caption{Average Binder cumulant $g_\mathrm{av}$ vs.\ temperature $T$ for anisotropic
interactions with $\Delta J=0.005$ and several system sizes $L$. $p=1/4$, $J_1=-J_2=1$.
The data are averages over 10,000 to 20,000 disorder configurations. The statistical
errors are smaller than the symbol size.}
\label{fig:0005_gav}
\end{figure}

To investigate whether or not the systems feature a phase transition into a long-range
ordered stripe phase, we analyze the average Binder cumulant $g_\mathrm{av}$.
For all $\Delta J \ge 0.005$, we find that the stripe Binder
cumulant curves for different system sizes $L$ cross at a nonzero temperature, indicating
the existence of the phase transition.
Examples of the average Binder cumulant data are presented in Figs.\ \ref{fig:02_gav} and \ref{fig:0005_gav}.
The global Binder cumulant behaves analogously. The curves for $\Delta J =0.2$ (Fig.\ \ref{fig:02_gav})
display a nearly perfect crossing for all considered system sizes, demonstrating that corrections
to scaling are weak. For  $\Delta J =0.005$ (Fig.\ \ref{fig:0005_gav}), in contrast,
the curves for smaller system sizes ($L < 64$) do not cross and resemble the isotropic $\Delta J =0$ case.
The curves for larger systems cross but the crossing temperature of consecutive curves shifts
systematically to higher values with increasing $L$. This indicates that the data for the studied
system sizes have not quite reached the asymptotic critical regime.

The fact that the Binder cumulant curves for smaller sizes do not cross for weak anisotropy
is readily understood by comparing the random field energy at a given system size with the energy
gain for horizontal stripes due to $\Delta J$. According to eq.\ (\ref{eq:E_RF}), the typical
energy gain due to aligning a domain of size $L$ with the local random fields is
$h_\mathrm{eff} L = \sqrt{2} p J_1 L$ whereas the anisotropy favors horizontal
stripes by the energy $\Delta J L^2$. A weak anisotropy can thus only suppress vertical domains
of sizes larger than $L_{\Delta J} \approx \sqrt{2} p J_1/ \Delta J$
\footnote{Note that this is a very rough estimate as it assumes
compact domains which only holds on scales below $L_0$. On larger length scales the cluster
structure is fractal, see Sec.\ \ref{subsec:domain}.}.
For $\Delta J =0.005$, this
estimate gives $L_{\Delta J} \approx 70$ in agreement with the observation that crossings start to appear
for $L \ge 64$. For $\Delta J = 0.002$, the smallest domain that the anisotropy can flip
has a size of about $L \approx 175$. As our system sizes are restricted to $L \le 192$ , this explains why
we do not observe clear crossings of the Binder cumulant curves for $\Delta J = 0.002$.
In other words, identifying the phase transition for $\Delta J \le 0.002$ requires simulations
of significantly larger systems.

We now analyze how the transition temperature $T_c$ into the stripe-ordered phase varies with the interaction
anisotropy $\Delta J$. To this end, we determine the crossing temperature for each $\Delta J$ value. This is
unambiguous for the larger $\Delta J$ for which the crossing is ``sharp'', i.e., the curves all cross
at the same temperature within their statistical errors. For the smaller $\Delta J$, where the crossing
shifts with increasing $L$, we estimate $T_c$ from the crossing of the largest two system sizes
\footnote{A systematic extrapolation of the crossing temperature to infinite system size, as performed,
e.g., in Ref.\ \cite{ZWNHV15}, would require significantly lower statistical errors and is thus beyond
our current computational capabilities.}.

\begin{figure}
\includegraphics[width=\columnwidth]{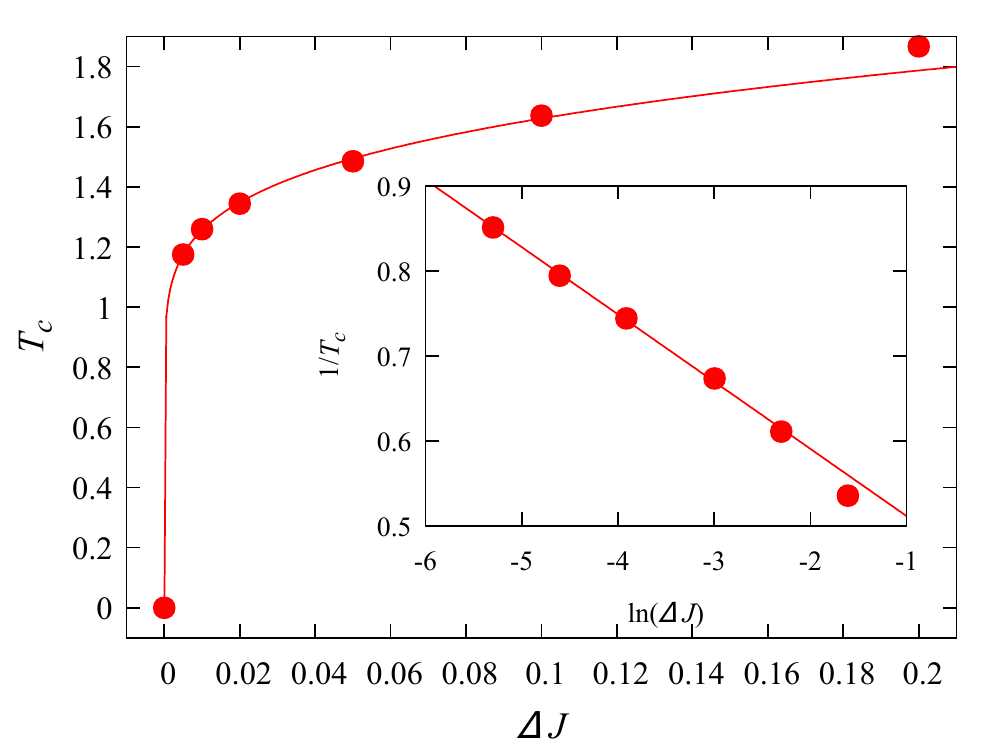}
\caption{Transition temperature $T_c$ into the long-range stripe ordered phase
vs.\ interaction anisotropy $\Delta J$ for $p=1/4$, $J_1=-J_2=1$. The solid line is a fit
of the data for $\Delta J < 0.2$ with the logarithmic dependence (\ref{eq:Tc})
yielding $1/T_c = -0.0791 \ln(\Delta J)+ 0.432$.
Inset: Data replotted as $1/T_c$ vs.\ $\ln \Delta J$ such that (\ref{eq:Tc}) leads to a straight line.}
\label{fig:Tc}
\end{figure}

Figure \ref{fig:Tc} presents the resulting dependence of $T_c$ on $\Delta J$.
The data show that $T_c$ rises very rapidly as $\Delta J$ increases from zero implying
that a small global anisotropy is sufficient to stabilize a robust stripe phase.
The figure also demonstrates that $T_c$ follows the logarithmic dependence (\ref{eq:Tc})
on $\Delta J$ predicted in Sec.\ \ref{subsec:transition} for all $\Delta J \le 0.1$.

It is interesting to compare the critical temperatures in Fig.\ \ref{fig:Tc} with
the corresponding value $T_{c0} \approx 2.08$ \cite{KalzHoneckerMoliner11} for the
undiluted isotropic system at the same parameter values ($J_1=-J_2=1$). Our simulations
show that a weak anisotropy of $\Delta J=0.005$ already produces a $T_c$ of more than
half of the undiluted value. Moreover, a large part of the reduction can be attributed
to the random-mass effects of the dilution in our system and not the random-field physics.
Thus, a better comparison may be the diluted system with anticorrelated impurities
studied in Ref.\  \cite{KunwarSenVojtaNarayanan18}. In that system, the random-field
physics is completely eliminated by the vacancy anticorrelations. Its
critical temperature of $T_c \approx 1.17$  (for $p=1/4$ and $J_1=-J_2=1$) is comparable
to the critical temperatures in Fig.\ \ref{fig:Tc} for anisotropies $\Delta J$ that have
largely suppressed the effects of the random-field disorder.

\subsection{Critical behavior}
\label{subsec:res_critical}

According to the discussion in Sec.\ \ref{subsec:transition}, we expect the transition into
the long-range stripe-ordered phase to be continuous and to belong to the two-dimensional
disordered Ising universality class. A perturbative renormalization group approach
\cite{DotsenkoDotsenko83,Shalaev84,Shankar87} predicts its critical behavior to be controlled
by the clean Ising fixed point while the disorder gives rise to universal logarithmic corrections
to scaling. This leads to the following finite-size scaling behavior \cite{MazzeoKuhn99,HTPV08,KennaRuizLorenzo08}.
The specific heat at the critical temperature diverges as
\begin{equation}
C \sim \ln \ln L~
\label{eq:C_lnln}
\end{equation}
with system size $L$.
The order parameter and its susceptibility at $T_c$ behave as
\begin{eqnarray}
\psi  &\sim& L^{-\beta/\nu}\, [1+O(1/\ln L)]~,
\label{eq:M_ln}\\
\chi_s &\sim& L^{\gamma/\nu}\,[1+O(1/\ln L)]~,
\label{eq:chi_ln}
\end{eqnarray}
with $\beta/\nu=1/8$ and  $\gamma/\nu=7/4$ as in the clean two-dimensional
Ising model. Any quantity $R$ of scale dimension zero
(such as the Binder cumulants $g_\mathrm{av}$ and $g_\mathrm{gl}$)
and its temperature derivative scale as
\begin{eqnarray}
R &=& R^\ast + O(1/\ln L)~,
\label{eq:R_ln}\\
dR/dT  &\sim& L^{1/\nu}(\ln L)^{-1/2}\, [1+O(1/\ln L)]
\label{eq:dRdT_ln}
\end{eqnarray}
with the clean Ising value $\nu=1$.

Identifying logarithmic corrections in numerical simulations and distinguishing them
from power laws with small exponents requires high-quality data over a significant system-size
range. Here, we therefore focus on $\Delta J = 0.2$ for which the system reaches
the asymptotic critical regime for smaller $L$ than for weaker anisotropies (see Figs.\
\ref{fig:02_gav} and \ref{fig:0005_gav}). We also simulate more disorder configurations
for $\Delta J =0.2$ than for the other $\Delta J$ to further reduce the statistical errors.

\begin{figure}
\includegraphics[width=\columnwidth]{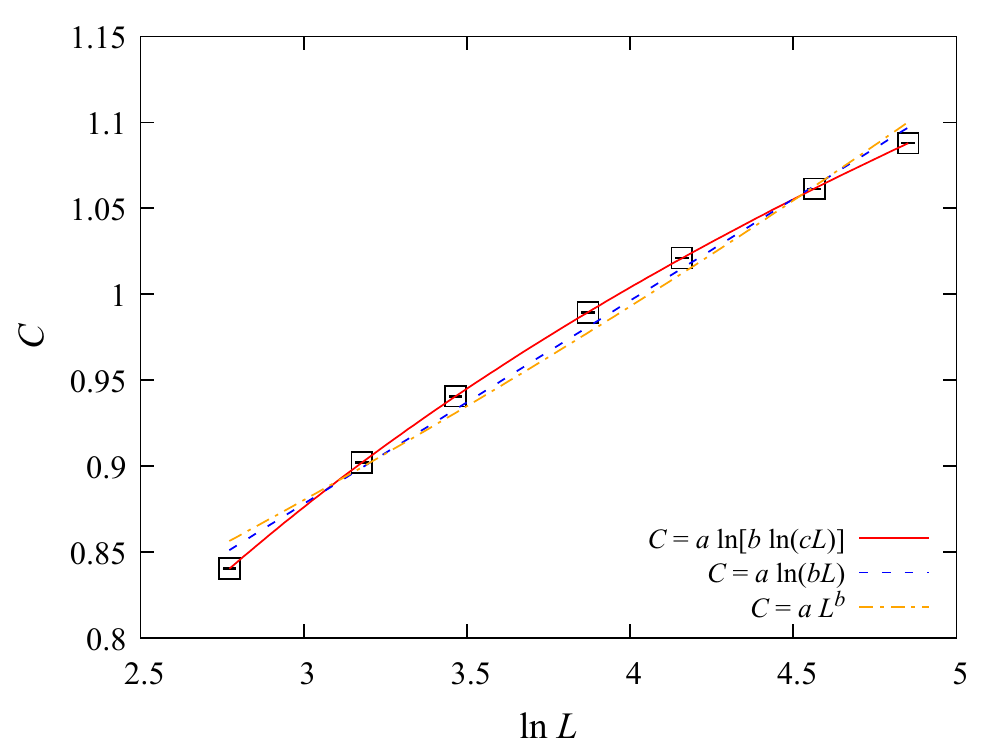}
\caption{Semilog plot of the specific heat $C$ vs.\ system size $L$ at the critical temperature $T_c=1.8670$ for
$\Delta J =0.2$, $J_1=-J_2=1$, $p=1/4$. The data are averages over 30,000 to 100,000 disorder
configurations. The resulting statistical errors are much smaller than the symbol size.
The solid line represents a fit with $C=a \ln[b \ln(cL)]$. The dashed and
dash-dotted lines represent a simple logarithmic fit $C=a \ln(bL)$ and a power-law fit
$C=a\, L^b$, respectively.}
\label{fig:Cv}
\end{figure}

To test the theoretical predictions (\ref{eq:C_lnln}) to (\ref{eq:dRdT_ln}), we analyze
the system-size dependence of $C$, $\psi$, $\chi_s$, and $dg_\mathrm{av}/dT$ at the
critical temperature $T_c=1.8670$. (We use polynomial interpolations in $T$ to determine
these values from the simulation data.) Figure \ref{fig:Cv} presents a semilogarithmic plot
of the specific heat $C$ vs.\ the system size $L$.
The figure clearly shows that the specific heat grows slower than logarithmic with $L$. It can be fitted well with
the double-logarithmic form $a \ln[b \ln(cL)]$ suggested by eq.\ (\ref{eq:C_lnln}), giving a reduced error sum $\bar \chi^2$
below unity \footnote{ For fitting $n$ data points $(x_i,y_i)$ to a function $f(x)$ containing $q$ fit parameters, the reduced error
sum is defined as
\begin{equation}
\bar \chi^2 = \frac 1 {n-q} \sum_i \frac{(y_i - f(x_i))^2}{\sigma_i^2}
\end{equation}
where $\sigma_i^2$ is the variance of $y_i$. The fits are considered to be of good quality if $\bar \chi^2 \lessapprox 2$.}.
In contrast, both a simple logarithmic fit $C=a\ln(bL)$ and a power-law fit $C=a\, L^b$ lead to unacceptably large
reduced $\bar \chi^2$ values of about 800 and 1600, respectively.

\begin{figure}
\includegraphics[width=\columnwidth]{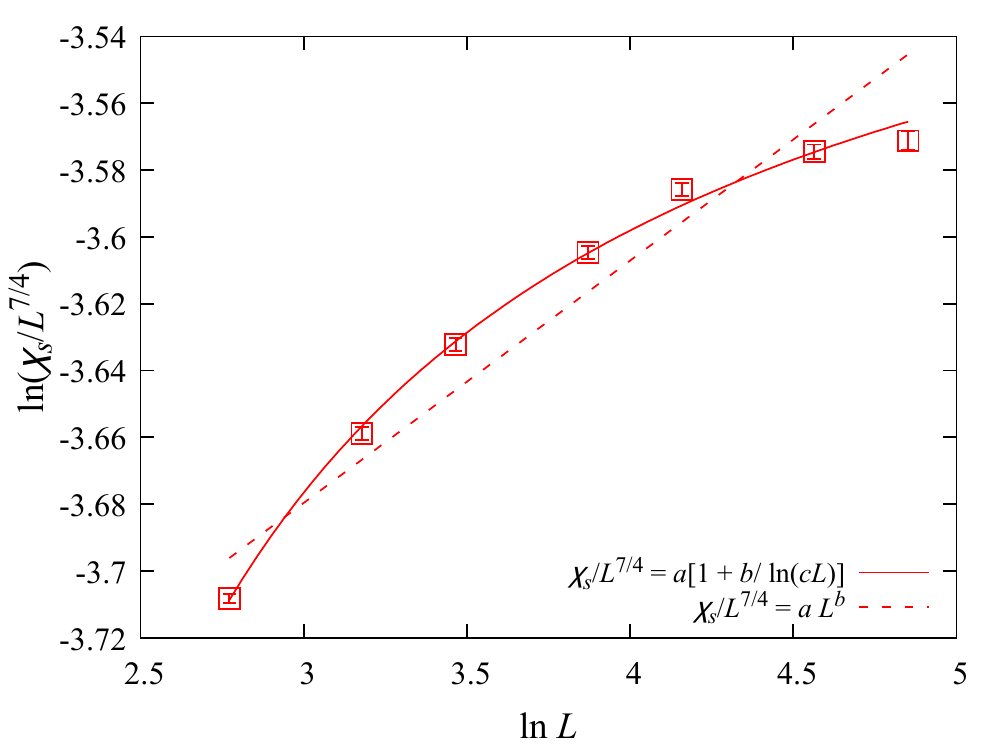}
\caption{Double logarithmic plot of $\chi_s L^{-7/4}$ vs.\ system size $L$ at the critical temperature $T_c=1.8670$ for
$\Delta J =0.2$, $J_1=-J_2=1$, $p=1/4$. The data are averages over 30,000 to 100,000 disorder
configurations. The solid line represents a fit with $a[1+b/\ln(cL)]$. The dashed line represents a
simple power-law fit with the functional form $a\, L^b$.}
\label{fig:chi}
\end{figure}

To test the predicted behavior (\ref{eq:chi_ln}) of the stripe susceptibility, we divide out the clean
Ising power law and plot $\chi_s L^{-7/4}$ vs.\ $L$ in Fig.\ \ref{fig:chi}.
The figure demonstrates that
$\chi_s L^{-7/4}$ increases more slowly than a power law with $L$. The data can be fitted reasonably well
with the form $a[1+b/\ln(cL)]$, yielding a reduced error sum of $\bar \chi^2 \approx 2.9$. (The reduced error sum
drops to about 1.3 if the smallest system size, $L=16$, is discarded.) A power-law fit
produces a unacceptably large $\bar \chi^2$ of about 60. The stripe order parameter can be treated
analogously, i.e., by analyzing $\psi L^{1/8}$. However, the corrections to the clean Ising behavior
for $\psi$ are much weaker than those for $\chi_s$, they only lead to a relative variation of $\psi L^{1/8}$
by about 1\% over the size range from $L=16$ to 128. Within the given statistical errors, both (\ref{eq:M_ln})
and a power law $\psi \sim L^{-\beta/\nu}$ with $\beta/\nu \approx 0.120$ fit the data.

Finally, we analyze the system-size dependence of the slopes $dg_\mathrm{av}/dT$ of the Binder cumulant
curves at criticality.
Within the statistical errors of our data and the uncertainty of $T_c$, we cannot discriminate
between Eq.\ (\ref{eq:dRdT_ln}) and simple power law $dg_\mathrm{av}/dT \sim L^{1/\nu}$
(which gives $\nu \approx 1.12$). Both functional forms fit the data reasonably well.

Taken together, the analyses of  $C$, $\psi$, $\chi_s$, and $dg_\mathrm{av}/dT$ provide strong evidence
for the critical behavior to belong to the two-dimensional disordered Ising universality class,
characterized by the clean Ising exponents with universal logarithmic corrections. To confirm that this behavior
also holds for smaller anisotropies, we have studied the system size dependence of the specific heat at criticality
for the other simulated $\Delta J$ values. For all $\Delta J > 0.01$, the specific heat data can be fitted well
with the double logarithmic form (\ref{eq:C_lnln}), giving  reduced error sums around unity. Even for the smallest
$\Delta J = 0.01$ and 0.005, the double logarithmic form fits much better than a simple logarithmic dependence or a power law.
However, the fit quality is noticeably worse ($\bar \chi^2 \approx 3$ and 6, respectively). This can be attributed to the fact that
the systems with $\Delta J \le 0.01$
have not reached the asymptotic critical regime in the size range $L=16$ to 128 (see Fig.\ \ref{fig:0005_gav}).

\section{Conclusion}
\label{sec:conclusion}

To summarize, we have investigated the combined influence of spinless impurities and a spatial interaction anisotropy
on the low-temperature stripe phase in the frustrated square-lattice $J_1$-$J_2$ Ising model.
The impurities reduce the effective interaction strength and thus create random-mass disorder.
They also locally break the $C_4$ rotation symmetry of the lattice, and thus create effective random fields
coupling to the nematic order parameter that distinguishes the two possible stripe directions.
In the absence of a global anisotropy, these random fields destroy the stripe phase
via domain formation.

A global interaction anisotropy that explicitly breaks the $C_4$ lattice symmetry competes with the random fields
and restores the stripe phase at sufficiently low temperatures. By combining percolation theory and results about the
domain structure of a biased random-field Ising model, we have predicted that the transition temperature $T_c$
into the stripe phase varies as  $T_c \sim 1/|\ln (\Delta J)|$ with the interaction anisotropy $\Delta J$. This means
very small $\Delta J$ are sufficient to restore a robust stripe phase.

We have also studied the resulting phase transition into the stripe phase. Our Monte Carlo results provide strong numerical
evidence for the transition to be continuous and to belong to the disordered two-dimensional Ising universality class
which is characterized by the clean Ising exponents and universal logarithmic corrections.

Our explicit calculations have implemented the global anisotropy via a difference between the
nearest-neighbor interactions in the two lattice directions. Other sources of global anisotropies that
break the symmetry between the two stripe directions are expected to have analogous effects. For example,
a global anisotropy in the impurity distribution that favors impurity pairs on, say, horizontal
nearest neighbor sites over pairs on vertical nearest neighbor sites introduces a bias into
the random field distribution. Horizontal stripe domains thus proliferate and form a massive spanning
cluster, just as in our case.

Let us also comment on the possibility of a nematic phase. In the absence of a global anisotropy,
($\Delta J = 0$), the phase transition between the paramagnetic high-temperature phase and the
stripe low-temperature phase, if any, could in principle split into two separate transitions,
the first breaking the $C_4$ lattice symmetry, producing nematic order, and the second breaking the
Ising spin symmetry. In the clean $J_1$-$J_2$ Ising model, a nematic phase has not been observed,
and same holds for the diluted model studied in Ref.\ \cite{KunwarSenVojtaNarayanan18} in which
the random-field physics is suppressed by impurity anti-correlations. The $J_1$-$J_2$ Heisenberg
model, in contrast, hosts a nematic phase \cite{ChandraColemanLarkin90}.
We emphasize that a nematic phase transition cannot occur in principle in the presence of
of a nonzero anisotropy $\Delta J \ne 0$. The anisotropy breaks the $C_4$ lattice symmetry
explicitly, spontaneous breaking of this symmetry is thus impossible \footnote{This argument does not preclude
more complicated scenarios such as a meta-nematic transition in analogy to a meta-magnetic transition
in a ferromagnet.}.

Our results have demonstrated that the random-field effects generated by spinless impurities
(and, by analogy, bond dilution or other types of quenched randomness)
on an order parameter that breaks a real-space symmetry are very sensitive to weak global spatial
anisotropies. This may complicate the experimental observation of the random-field physics, for
example if the samples feature residual strain. A systematic variation of the anisotropy to test the
predictions of the present paper may be achieved, e.g., by applying uniaxial pressure.

We note that the interplay and feedback between the random-field induced domain formation
and the magnetic degrees of freedom leads to enhanced fluctuations and slow dynamics even in the absence
of a global anisotropy, as was recently demonstrated by mapping the $J_1$-$J_2$ Hamiltonian on an
Ashkin-Teller model in a random Baxter field \cite{MeeseVojtaFernandes21}.

It is interesting to compare our results to those for the square-lattice $J_1$-$J_2$ Heisenberg model.
Even though magnetic long-range order at nonzero temperatures is impossible in the Heisenberg case
due to the Mermin-Wagner theorem \cite{MerminWagner66}, the clean $J_1$-$J_2$ Heisenberg model
features vestigial nematic order \cite{ChandraColemanLarkin90} associated with the unrealized
stripe phase (for $|J_2| > J_1/2$). Fyodorov and Shender \cite{FyodorovShender91} argued that
random bond dilution creates random fields for the nematic order just as in the Ising case,
destroying the nematic phase. Recently, Miranda et al.\ \cite{Mirandaetal21} demonstrated that this conclusion
holds generically for both bond disorder and site vacancies.
As a result, the system is a nontrivial paramagnet for nonzero temperatures, and a spin-vortex-crystal
glass for zero temperature and weak disorder \cite{Mirandaetal21}.

Impurity-induced random fields also emerge in three-dimensional frustrated magnets. For example,
in XY pyrochlore magnets, they have recently been shown to destroy long-range order beyond a
critical disorder strength,  leading to the formation of a cluster-glass state \cite{AndradeHoyosRachelVojta18}.

The use of strain to manipulate and ``engineer'' phases and properties of many-particle systems has recently
attracted considerable attention. For instance, it was realized that strain can lift
the degeneracy of the ground state manifold of a frustrated Heisenberg antiferromagnet on a Kagome lattice,
tuning the system through a sequence of unconventional phases \cite{NaygaVojtaM21}. Our results can
be understood as an example of using strain engineering to restore the stripe phase.

\acknowledgments

The work in Missouri has been supported in part by the National Science Foundation under Grant Nos.\  DMR-1828489 and
OAC-1919789. The simulations were performed on the Pegasus and Foundry clusters at Missouri S\&T.
RN acknowledges funding from the Center for Quantum Information Theory in Matter and Spacetime, IIT Madras,
and from the Department of Science and Technology, Govt. of India, under Grant No. DST/ICPS/QuST/Theme-3/2019/Q69.
We also thank Rafael Fernandes and Joe Meese for helpful discussions.

\bibliographystyle{apsrev4-2}
\bibliography{../00Bibtex/rareregions}

\end{document}